# Synthetic multi-focus optical-resolution photoacoustic microscope for large volumetric imaging

Xianlin Song [a, #, *], Jianshuang Wei [b, c, #], Lingfang Song [d]

[a] School of Information Engineering, Nanchang University, Nanchang, China;

[b] Britton Chance Center for Biomedical Photonics, Wuhan National Laboratory for Optoelectronics-Huazhong University of Science and Technology, Wuhan, China;

[c] Moe Key Laboratory of Biomedical Photonics of Ministry of Education, Department of Biomedical Engineering, Huazhong University of Science and Technology, Wuhan 430074, China;

[d] Nanchang Normal University, Nanchang 330031, China;

[#] equally contributed to this work

*songxianlin@ncu.edu.cn

**Abstract**：Photoacoustic microscopy is becoming an important tool for the biomedical research. It has been widely used in biological researches, such as structural imaging of vasculature, brain structural and functional imaging, and tumor detection. The conventional optical-resolution photoacoustic microscopy (OR- PAM) employs focused gaussian beam to achieve high lateral resolution by a microscope objective with high numerical apertures. Since the focused gaussian beam only has narrow depth range in focus, little detail in depth direction can be revealed. Here, we developed a synthetic multi-focus optical-resolution photoacoustic microscope using multi-scale weighted gradient-based fusion. Based on the saliency of the image structure, a gradient-based multi-focus image fusion method is used, and a multi-scale method is used to determine the gradient weights. We pay special attention to a dual-scale scheme, which effectively solves the fusion problem caused by anisotropic blur and registration error. First, the structure-based large-scale focus measurement method is used to reduce the effect of anisotropic blur and registration error on the detection of the focus area, and then the gradient weights near the edge wave are used by applying the small-scale focus measure. Simulation was performed to test the performance of our method, different focused images were used to verify the feasibility of the method. Performance of our method was analyzed by calculating Entropy, Mean Square Error (MSE) and Edge strength. The result of simulation shown that this method can extend the depth of field of PAM two times without the sacrifice of lateral resolution. And the in vivo imaging of the zebra fish further demonstrates the feasibility of our method.

## 1 Introduction

Photoacoustic imaging is a new medical imaging method based on the photoacoustic effect[1]. The photoacoustic effect was proposed in 1880 by Bell. A laser pulse irradiates the sample, the sample absorbs pulse energy, then the temperature rises instantaneously, and adiabatic expansion occurs, which then generates ultrasonic signal. Because the ultrasonic signal is excited by laser pulses, it is called photoacoustic signal. The photoacoustic effect can be successfully applied to biological tissue imaging because of the large differences in the absorption of laser pulses in different biological tissues[2-4]. The photoacoustic signal received by the ultrasonic probe contains the absorption characteristics of biological tissues. The optical absorption distribution can be reconstructed by the reconstruction algorithm.

As a promising tool for biomedical research, optical-resolution photoacoustic microscopy (OR- PAM) is a noninvasive biomedical imaging technique with high-resolution[5]. Various applications has been implemented, which include structural, functional and molecular imaging of tissues[6-8]. The conventional OR-PAM employs focused gaussian beam to achieve high lateral resolution by a microscope objective with high numerical apertures, which can achieve micron- to submicron-sized focal spot. Since the focused gaussian beam only has narrow depth range in focus, little detail in depth direction can be revealed[9]. To address this issue, many methods are developed to extend the depth of field (DoF). The non-diffracting properties of bessel beam make it own large depth of field compare with gaussian beam[10], while the artifacts introduced by the side lobes of the Bessel beam should be suppressd by non-linear method. Electrically tunable lens (ETL) has also been introduced in OR-PAM[11]. This resulted in a focus-shifting time of about 15 ms. It is fast enough for pulsed lasers with a repetition rate of tens of hertz, while being quite slow for those lasers with repetition rate of kilos to hundreds of kilohertz, which has been widely used in the OR-PAM. Tunable acoustic gradient lens has also been introduced into OR-PAM[12, 13], but it needs special synchronization circuits to control the laser, which made system more complicated. Depth scanning by using motorized stage is commonly used as it is the most convenient approach[14, 15]. However, this method limits the volumetric imaging speed with its slow mechanical adjustment, and it is a challenge that the images obtained by scanning at different depth fuse into a clear image.

In this manuscript, we reported a synthetic multi-focus optical-resolution photoacoustic microscope using multi-scale weighted gradient-based fusion. Based on the saliency of the image structure, a gradient-based multi-focus image fusion method is used, and a multi-scale method is used to determine the gradient weights. We pay special attention to a dual-scale scheme, which effectively solves the fusion problem caused by anisotropic blur and registration error. First, the structure-based large-scale focus measurement method is used to reduce the effect of anisotropic blur and registration error on the detection of the focus area, and then the gradient weights near the edge wave are used by applying the small-scale focus measure. Simulation was performed to test the performance of our method, different focused images were used to verify the feasibility of the method. Performance of our method was analyzed by calculating Entropy, Mean Square Error (MSE) and Edge strength. The result of simulation shown that this method can extend the depth of field of PAM two times without the sacrifice of lateral resolution. And the in vivo imaging of the zebra fish further demonstrates the feasibility of our method.

## 2 METHOD

**2.1 Multi-scale weighted gradient-based fusion (MWGF)**

The gradient-based weighted fusion method can effectively identify the most important local structure in the input image and render it into the fusion image, which is suitable for a variety of fusion applications, such as multi-spectral fusion, multi-focus fusion and medical image fusion. In order to more effectively fuse the focus areas of multi-focus images, many fusion methods based on focus measures have been proposed. A common method for multi-focus image fusion using focus measurement is to divide the input image into blocks, calculate the sharpness of the blocks, and then select the blocks with higher sharpness to reconstruct the fused image. However, the efficiency of this method largely depends on the size of the block, that is, the scale of the focal point measurement. Due to anisotropic blur or misregistration, a large number of blocks in the out-of-focus area may be sharper than the blocks in the focus area. This will result in the selection of an out-of-focus area.

In order to solve the above problems, we use the weighted gradient fusion method under multi-scale[16], this method consists of 5 main steps. First, given multi-focus images $I_n(x, y), n = 1,...,N$, calculate the gradient of each source image. The max amplitude projection (MAP) images (source images) were acquired at different depth by homemade OR-PAM[13].

Then, the large-scale focus measure is used to roughly detect the focus area of each input image. In the binary initial detection result, the smooth area in the largest connected area that is easy to be misdetected can be regarded as a part of the area, and the filling operation can be performed. Optionally, the binary initial detection result can also be pre-processed by closing operation to fill the small gaps between different detection areas.

Third, defining an unknown area near the boundary of the focus area to obtain a clear focus area and an off-focus area. Setting the gradient weight of the focus area and the out-of-focus area to 1 and 0, respectively.

Use the small-scale focus measure to determine the gradient weights by combining multi-scale information in the unknown region. Finally, use gradient-based weighted fusion method to construct fusion image.

## 3 RESULTS

### 3.1 Simulation study

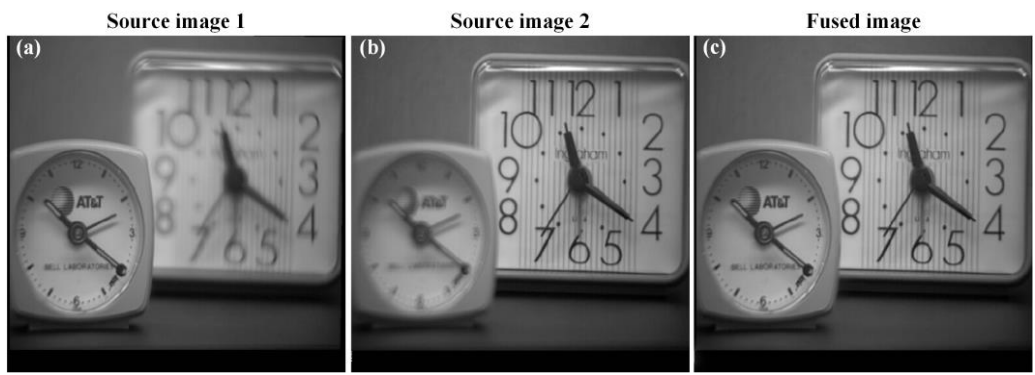

Figure 1. Simulation. (a) Source image 1, the front clock is focused, and the back clock is out of focus. (b) Source image 2, the front clock is out of focus, and the back clock is focused. (c) Fused Image, both the front clock and the back clock are focused, more details can be resolved along the depth.

Different focused images were used to verify the feasibility of the method. As shown in Figure 1, Figure 1 (a) and Figure 1 (b) are the same images with different focus. Figure 1 (a) front clock is focused, back clock is out of focus. Figure 1 (b) front clock is out of focus, back clock is focused. Figure 1 (c) is obtained after using the fusion rules, both the front clock and the back clock are focused, more details can be resolved along the depth. The fused image contains more informative compare to the individual source images. Therefore, computed extended depth of field was achieved by using our method. Performance of our method was analyzed by calculating Entropy, Average gradient, Mean Square Error (MSE) and Edge strength, as shown in Table 1.

Table 1. Performance measure of image fusion.

|  | Entropy | Average gradient | Standard deviation | Edge strength |
|---|---|---|---|---|
| **Image 1** | 7.3141 | 2.0756 | 51.8651 | 20.6003 |
| **Image 2** | 7.3706 | 2.4499 | 51.6847 | 25.0263 |
| **Fused image** | 7.3989 | 2.6662 | 52.3305 | 27.3982 |

Entropy is a major indicator of the richness of information in an image, which indicates the average amount of information contained in an image. The fused image has bigger entropy value (7.3989) compare to the source image 1 (7.3141) and source image 2 (7.3706), respectively. The average gradient represents the contrast of details of image. The average gradient is also called the sharpness of the image, reflecting the contrast and texture transformation characteristics of detail. The average gradient of fused image is 2.6662, while, the average gradient of image1 and image 2 is 2.0756 and 2.4499, respectively.

Standard deviation is a statistical concept that expresses the degree of dispersion, and it reflects the distribution of gray values of an image. The magnitude of the standard deviation is positively related to the quality of the image. The larger the standard deviation, the more scattered the grayscale distribution of the image, the more useful information the image has. The standard deviation of fused image is 52.3305, higher than that of image1 (51.8651) and image 2 (51.6847), respectively. The edge strength was also used to evaluate the performance of our method, the edge strength of fused image is 27.3982, while, the edge strength of image1 and image 2 is 20.6003 and 25.0263, respectively. The quality of the fused image is significantly better than that of a single source image, and it contains more details, which indicates a larger (~ two times) depth of field the fused image has.

We use this method to compare with other image fusion methods, the results is shown in Figure 2. Figures 2(a)-2(f) are fused images by using LAP, DWT, PCA, GRA, FSD and MWGF, respectively. It can be seen that compared to other fusion methods, the fused image obtained by the MWGF method is clearer and the image contrast is the best.

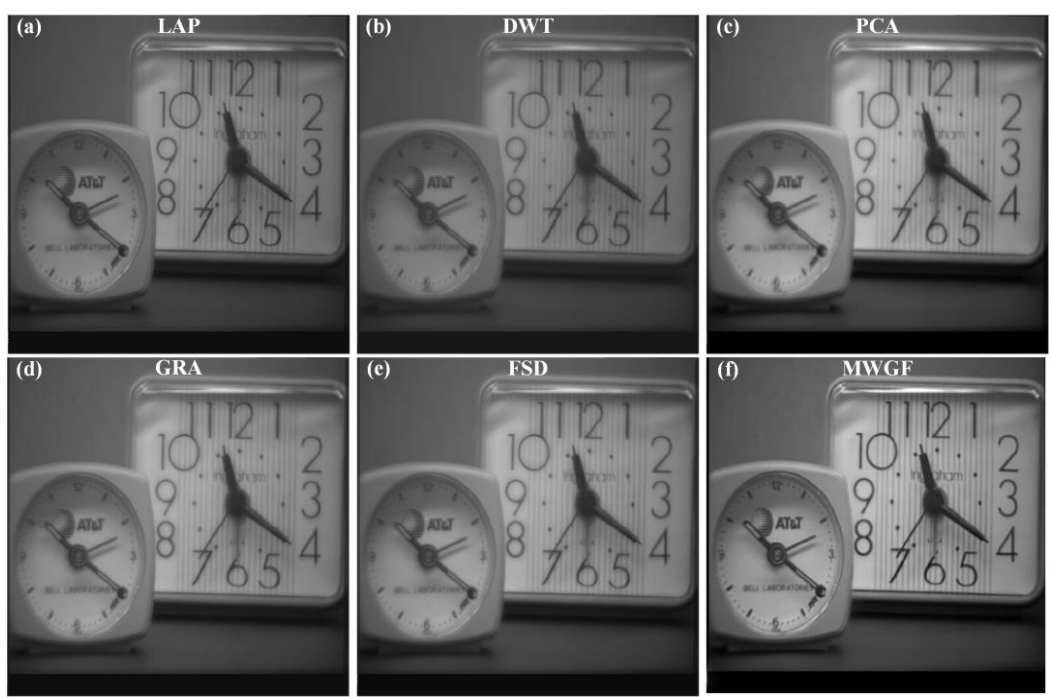

Figure 2. Fusion results of the "Clock" source images. (a)-(f) are fused images by using LAP, DWT, PCA, GRA, FSD and MWGF, respectively.

### 3.2 *In vivo* imaging of zebrafish

To fully demonstrate the performance of our method, A 3-day-old zebrafish (wild type) was chosen as the imaging object. Before the imaging, a culture dish was coated with a thin layer of low-melting point agarose (A-4018, Sigma-Aldrich), which was dissolved in 40 ℃ deionized water (1.2% w/v). When the temperature of the agarose dropped to 37 ℃, the zebrafish was placed and oriented in the culture dish so that it was in a side-lying, lightly covered with the melted liquid agarose, waiting for solidification. The temperature was kept around 25 ℃ during the imaging. All procedures were carried out in accordance with the Institutional Animal Care and Use Committee of Hubei Province.

A homemade single focus OR-PAM system with high resolution reached to micrometer as previously described[14], the DoF of the system is ~ 120 μm. First, we set the focal plane of the system to the surface of the zebra fish, a two-dimensional raster scanning with a step size of 3 μm is performed to acquire the depth-dependent PA signal (source PA image 1). Then, moving the focus down 100 μm, and perform two-dimensional raster scanning again to obtain source PA image 2.

As shown in Fig. 3, Fig. 3(a) and Fig. 3(b) are depth-coding max projection (MAP) images obtained when the focus located at z and z +100, respectively. Fig. 3(c) is the fused image of Fig. 3(a) and Fig. 3(b). Since the OR-PAM has a very limited depth of field, it can only clearly image structures with a limited depth, and the image will become blurred if it deviates slightly from the depth of field, as shown in Fig. 3(a) and Fig. 3(b). However, due to the use of image fusion methods, two virtual focal points (indicated by the white arrows) are generated, and the depth of field is also expanded to 2 times, which can obtain more depth information, as shown in Fig. 3(c).

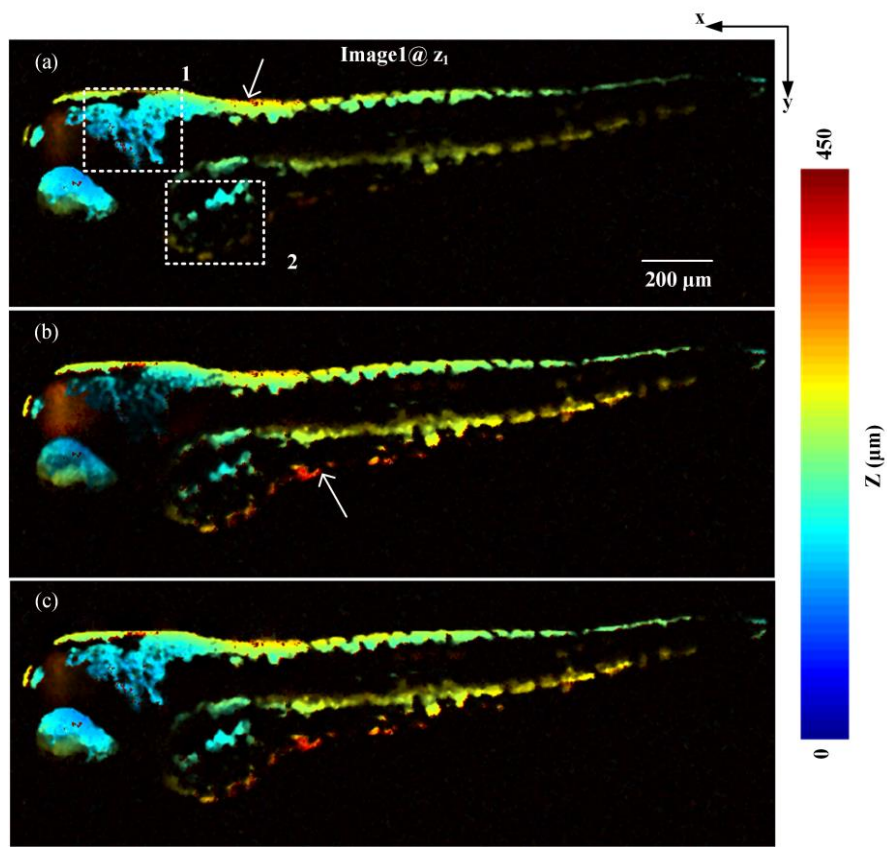

Figure 3. *In vivo* imaging of zebra fish. (a) and (b) are depth-coding max projection (MAP) images obtained when the focus located at z and z +100, respectively. (c) The fused image. The arrows in (a) and (b) indicate the approximate position of focus.

Figs. 4(a) - 4(c) are close-up MAP images of the areas indicated by the white rectangles 1 in Figs. 3(a) - 3(c), respectively; Figs. 4(d) - 4(f) are close-up MAP images of the areas indicated by the white rectangles 2 in Figs. 3(a) - 3(c), respectively. Since the narrow DoF of the OR-PAM, little details can be resolved, only part of the vessels are in focus in Fig. 3(a) and Fig. 3(b). While, the fusion rules enable the system to have a larger depth of field, more details from different depth can be distinguished. The pigments that are indicated by yellow arrows can be visualized in fused image, but fuzzy or missing in Fig. 3(a) and Fig. 3(b). This experiment demonstrates that the method we have proposed can form a synthetic multi-focus and thus extend the depth of field about two times.

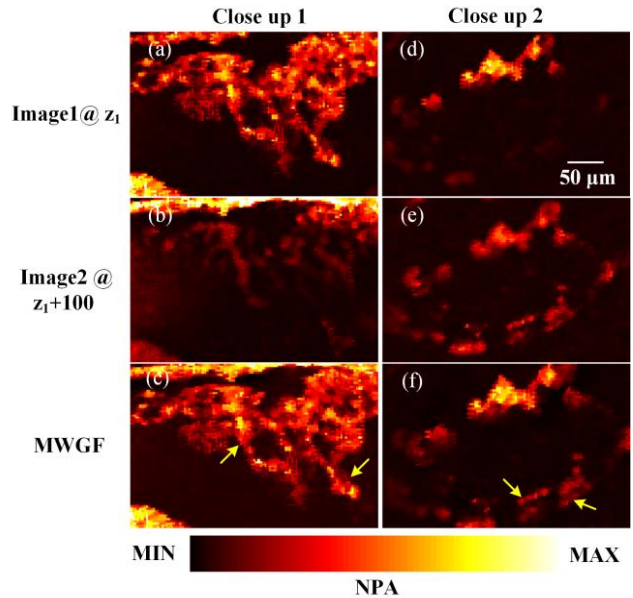

Figure 4. Close-up images. (a) - (c) are close-up images of the areas indicated by the rectangles 1 in Figures 3(a) - 3(c), respectively; (d) - (f) are close-up images of the areas indicated by the white dashed rectangles 2 in Figures 3(a) - 3(c), respectively. The yellow arrows in (c) and (f) denote the pigments which cannot be resolved in Figure 3(a) or Figure 3(b). NPA, normalized amplitude distribution.

## 4 Conclusion

In summary, by using multi-scale weighted gradient-based fusion into the PAM system, we developed a synthetic multi-focus optical-resolution photoacoustic microscope. A gradient-based multi-focus image fusion method is used, and a multi-scale method is used to determine the gradient weights. We pay special attention to a dual-scale scheme, which effectively solves the fusion problem caused by anisotropic blur and registration error. First, the structure-based large-scale focus measurement method is used to reduce the effect of anisotropic blur and registration error on the detection of the focus area, and then the gradient weights near the edge wave are used by applying the small-scale focus measure. Simulation and the in vivo imaging of zebra fish were performed to demonstrate that this method can extend the depth of field of PAM two times without the sacrifice of lateral resolution.